# Photo-Physical Characteristics of Boron Vacancy-Derived Defect Centers in Hexagonal Boron Nitride

Yifeng Chen[1,2] and Su Ying Quek[1,2,3,4]

1. Department of Physics, National University of Singapore, 2 Science Drive 3, Singapore 117551

2. Centre for Advanced 2D Materials, National University of Singapore, 6 Science Drive 2, Singapore 117546

3. NUS Graduate School Integrative Sciences and Engineering Programme, National University of Singapore, Singapore 117456

4. Department of Materials Science, National University of Singapore, 9 Engineering Drive 1, Singapore 117575

**Abstract:** Single photon emitter (SPE) sources are important building blocks for photonics-based quantum technologies. Recently, the highly bright and versatile SPEs from the two-dimensional insulator material hexagonal boron nitride (hBN) have attracted significant research interest. However, due to the variability of emitter species and properties, an exact correlation between the underlying atomistic structures and their photo-physical properties is still lacking. In this work, we study six boron vacancy-derived defect centers in hBN ($V_B^-$, $V_B$+H, $V_{B2}$, $V_BC_N$, $V_BC_N^-$, and $V_BC_NC_N$) using advanced first principles techniques, characterizing their quasiparticle defect levels, optical spectra and excitation energies, and magneto-resonance properties. These defects have been chosen because of their relatively low formation energies, and, importantly, because they are amenable to intentional creation under experimental conditions. We establish the correlation between the underlying defect atomic structure and their photo-physical properties, thus facilitating the identification of SPEs that have already been observed in experiments. Our results lead to clear insights into very recent experiments where hBN SPEs can be controlled intentionally. On the other hand, our results also serve as guidelines for the bottom-up design of defect emitter centers in hBN for target applications that require specific defect properties, such as emission in the telecom wavelength, optical addressability and high radiative decay rates. This work thus provides a comprehensive understanding of the photo-physical characteristics of $V_B$-derived defect emitting centers, aiding in their identification and manipulation for tailored applications.

**Introduction**

Single photon light sources are important building blocks for future photonics-based quantum sensing, quantum information processing and computing applications. The control and integration of these sources at the device and circuit levels are vital for realizing their full potential. In this context, solid state quantum emitters in wide band gap materials have attracted increasing attention.[1] Two-dimensional (2D) materials are emerging hosts of solid state SPEs[2-5] because of several advantages compared with conventional bulk hosts such as diamond and silicon carbide. It is easier to implant, fabricate and scale up on-demand emitting centers in 2D materials due to their layered structural nature. On the other hand, their ultra-thin nature also facilitates easy integration with photonic chips.[6] Recently, promising SPEs have been reported in 2D transition metal dichalcogenides,[7, 8] hexagonal boron nitride,[9] III-V monochalcogenides,[10] etc.

Hexagonal boron nitride is a wide bandgap insulator, with a bulk optical bandgap at 6.0 eV.[11] SPEs discovered in hBN can be broadly categorized into two groups. The first group is characterized by a ~2.1 eV (590 nm) emission zero phonon line (ZPL)[9] and several lower energy emitter families from the visible to near-infrared spectral range, e.g. 615, 650, 705, 812, 850 nm, etc.[12-15] The second group is characterized by emission in the ultraviolet range at around 4.1 eV.[16] Both of these categories involve deep level states within the hBN fundamental gap, with stable SPE properties at room temperature, and even up till ~800K.[17]

A recent exciting development in the field of SPEs in hBN is the discovery of optically-detected magnetic-resonance (ODMR) signals in hBN defect centers.[14, 18, 19] Specifically, the photoluminescence (PL) intensities are altered in the presence of microwave excitation that is resonant with the energy splitting between spin-sublevels in the triplet state, which are characterized by the zero field splitting (ZFS) parameters for the defect. The observation of ODMR signals in hBN defect centers implies that the corresponding defect centers can be initialized and read out by optical means, similar to the nitrogen-vacancy (NV$^-$) centers in diamond.[20] Such defect centers are thus very promising for photonics-based quantum technology applications.

Unlike the NV$^-$ centers, which have been studied in detail both theoretically and experimentally,[20, 21] the photo-physics of the experimentally observed SPEs in hBN is in general not well understood. A major difficulty is the lack of information on the detailed atomic structure of most SPEs in hBN. Uncovering the atomic structure of these SPEs will not only allow for further theoretical predictions on important figures of merit for the SPEs (e.g. intersystem crossing rates),[22] but can also provide guidance for the intentional creation of defect centers. As it is challenging to image defects in insulating hBN, spectroscopic measurements remain the primary method to probe the nature of SPEs in hBN.

ODMR measurements are particularly promising, as ODMR enables one to directly associate the measured PL energy with the zero-field splitting and Landé g-factors for the same defect, so that measured PL energies, ZFS and Landé g-factors serve as a viable fingerprint for each defect. However, spectroscopic experiments do not provide direct information on the atomic structure of the defects. *Ab initio* calculations of candidate defect centers are critical for the identification of SPEs in hBN. In recent work, an SPE center was identified to be the negatively charged boron vacancy defect $V_B^-$, based on ODMR signals and PL measurements in comparison with *ab initio* calculations.[23, 24] We note that ZFS parameters are non-zero only when there are more than one unpaired spin at the defect. Another commonly measured quantity, obtained with electron paramagnetic resonance (EPR) measurements, is the hyperfine coupling tensor, which gives information about the atomic species but not on the specific atomic structure. Although EPR signals typically arise from an ensemble of defects, they nonetheless provide information that can potentially be useful in identifying the SPEs in hBN. For example, rotation of the sample with respect to the applied magnetic field can provide angle-dependent EPR measurements that provide information on the local symmetry of the defect site.[14]

In this work, we perform state-of-the-art first principles calculations to study the photo-physics of a number of boron vacancy ($V_B$)-derived defect centers in monolayer hBN. These defects include $V_B^-$, $V_B$+H, $V_{B2}$, $V_BC_N$, $V_BC_N^-$, and $V_BC_NC_N$. We predict their quasiparticle defect levels, optical spectra and excited state energies, as well as magneto-resonance (MR) paramaters. Our results can be directly compared with experimental observations, and are thus helpful for identifying SPEs that have already been observed in experiments. In particular, we find that $V_B$+H, $V_BC_N^-$ and $V_BC_NC_N$ defect centers are potential candidates for the ~ 2 eV emitters, widely observed in experiments under different sample preparation and characterization conditions.[9, 12, 13, 25-29] Based on the sample conditions and optical spectra, we predict that $V_B$+H is likely relevant for a recent observation of an in-plane Stark effect for a ~2 eV SPE,[29] $V_BC_N^-$ is responsible for the ~2 eV SPE in a recent carbon-implantation experiment,[27] and $V_BC_NC_N$ is a likely candidate for the ~2eV emission observed in recently synthesized carbon-enriched hBN.[28] The optical absorption spectra of these defects also display distinct anisotropic polarization-dependence. Our results show that these defects can be further distinguished by their MR parameters. We further predict that recent observations of position-controlled 2.8 eV SPE in electron-beam-irradiated hBN correspond to emission from $V_{B2}$ defects.[30] On the other hand, our results also serve as guidelines for the bottom-up design of defect emitter centers in hBN for specific applications. We predict the presence of spectrally-isolated lower-energy absorption peaks for $V_B$+H, $V_BC_N$ and $V_BC_NC_N$ defects, which renders them potentially useful in the telecom wavelengths. We find that $V_B^-$, $V_B$+H, $V_{B2}$ and $V_BC_N$ are all defects with triplet ground states and non-zero ZFS parameters, which are promising for the optical initialization of the defect centers. All the defects have reasonably large radiative decay rates which renders them useful as localized emitters. Among the defects, $V_{B2}$ has especially bright emission in the visible wavelengths. Finally, the out-of-plane dipole moment in $V_BC_NC_N$ suggests that the emission from $V_BC_NC_N$ can be tuned by a gate voltage. Our results thus provide a comprehensive understanding of the photo-physical characteristics of $V_B$-derived defect emitting centers, aiding in their identification and manipulation for tailored applications.

**Computational Approach**

We compute the defect levels from first principles using many-body perturbation theory within the GW approximation,[31] and predict the GW-BSE optical absorption spectra of the defects using periodic boundary conditions with a plane-wave basis set. These calculations represent state-of-the-art methods in predicting single- and two-particle properties for periodic systems.[32] Extensive studies have shown that GW calculations are able to predict quantitatively accurate quasiparticle levels in both bulk and low-dimensional materials, with no empirical fitting parameters.[33] The optical absorption spectra computed using the GW-BSE approach account for the effects of electron-hole interactions in excitons, which are particularly important in low-dimensional materials with reduced dielectric screening. Quantitatively accurate predictions of defect energy levels and optical spectra, as well as the relevant interband transition pairs, are a crucial first step to characterize and precisely manipulate the defect center-based SPEs in hBN. To our knowledge, there have only been a few GW-BSE calculations[22, 34, 35] on defects in hBN.

Defect centers involving dangling bonds or open-shell orbitals can sometimes be associated with many body electronic eigenstates with multi-reference character.[36] We therefore complement the GW-BSE calculations with density matrix renormalization group (DMRG) calculations on cluster models, allowing us to go beyond a single reference framework to predict excitation energies with potential multi-reference character.

We note that in this work, we have not considered the effects of phonons and structural relaxations in the excited states, which can cause the photoluminescence energies to be red-shifted from the vertical excitation energies or optical absorption peaks predicted here. According to the literature, these reorganization energies are on the order of a few hundred meVs.[3, 20, 24, 37-39]

Defect centers with unpaired electron spins have rich spin physics, which can be probed using ODMR and EPR measurements. Using density functional theory (DFT), we can accurately predict the magneto-resonance parameters, such as the *g*-tensor and zero field splitting (ZFS) parameters for spin-polarized defect centers, as well as the hyperfine coupling tensor. Since we know explicitly the underlying atomic structures of the defect centers in our calculations, a comparison with the experimentally measured values provides important fingerprinting information on the structural origins of the SPEs in experiments.

The GW-BSE calculations in this work are carried out using the BerkeleyGW software package,[31] with the DFT mean field starting point wavefunctions computed using the plane-wave pseudopotential code, Quantum-ESPRESSO.[40] The DMRG calculations are performed using the BLOCK code[41] with PySCF setups,[42] and spin-conserving vertical excitation energies are computed. The MR parameters are computed using all-electron DFT calculations from the ORCA code.[43] The DMRG calculations and the calculations of MR parameters are performed by modeling the defects in hBN using cluster models. For DMRG, two layers of rings are included around the defect centers (e.g., $B_{18}N_{18}H_{15}$ for $V_B^-$;[44] see Figure S1), while for the MR calculations, an additional layer of rings is included. Further details of these calculations can be found in the Methods Section.

**Results and Discussion**

We first describe briefly the results of our GW-BSE calculations on pristine hBN. Our GW calculations on bulk hBN show that it is an indirect band gap insulator, consistent with recent experimental findings.[11] The computed indirect band gap is 6.51 eV, while the direct band gap at **K** is 6.80 eV. For bulk hBN, we predict an optical onset energy of 5.62 eV with in-plane polarization, and 6.00 eV with out-of-plane polarization. For monolayer (ML) hBN, our GW band gap is 7.54 eV, direct at **K**, and the optical onset is 5.33 eV with in-plane polarization, corresponding to a large exciton binding energy of about 2.2 eV. These results are consistent with previous literature.[45]

The formation energies of various defects in hBN have been studied previously using DFT methods.[46] It was found that extrinsic impurity-induced defects, $C_B$, $C_N$, $O_N$, $V_B$+H, $V_B$+2H, $V_B$+3H, have lower formation energies than intrinsic defects. Kinetic factors during the material growth and processing steps also influence the nature and prevalence of different defects. For example, electron bombardment of hBN results in a larger concentration of boron vacancies than nitrogen vacancies in hBN.[47] Several of the SPE studies in hBN involve the intentional creation of defects using electron or ion irradiation.[13, 48, 49] Hence, in this work, we focus on a number of $V_B$-derived defect centers, such as $V_B^-$, $V_B$+H, and $V_{B2}$. We also study $V_B$-derived defect complexes incorporating $C_N$ antisites, namely $V_B C_N$, $V_B C_N^-$ and $V_B C_N C_N$, which are relevant for recent experiments where hBN was prepared in a carbon-rich environment.[27, 28] Such defects that can be intentionally created are of great interest for the ultimate goal of achieving on-demand bottom-up design of SPEs in hBN. The atomic structures of these defects are shown in Figure 1. We note that the supercell size used here is large enough to prevent interaction between defects. In particular, the defect levels are found to be flat, with a band dispersion of within 0.1 eV, except for the LUMO state of $V_BC_N$ which has a dispersion of 0.2 eV. $V_B^-$, $V_B$+H, $V_{B2}$ and $V_BC_N$ have spin triplet ground states (GS), which are promising for ODMR detection and photonics-based quantum information processing.

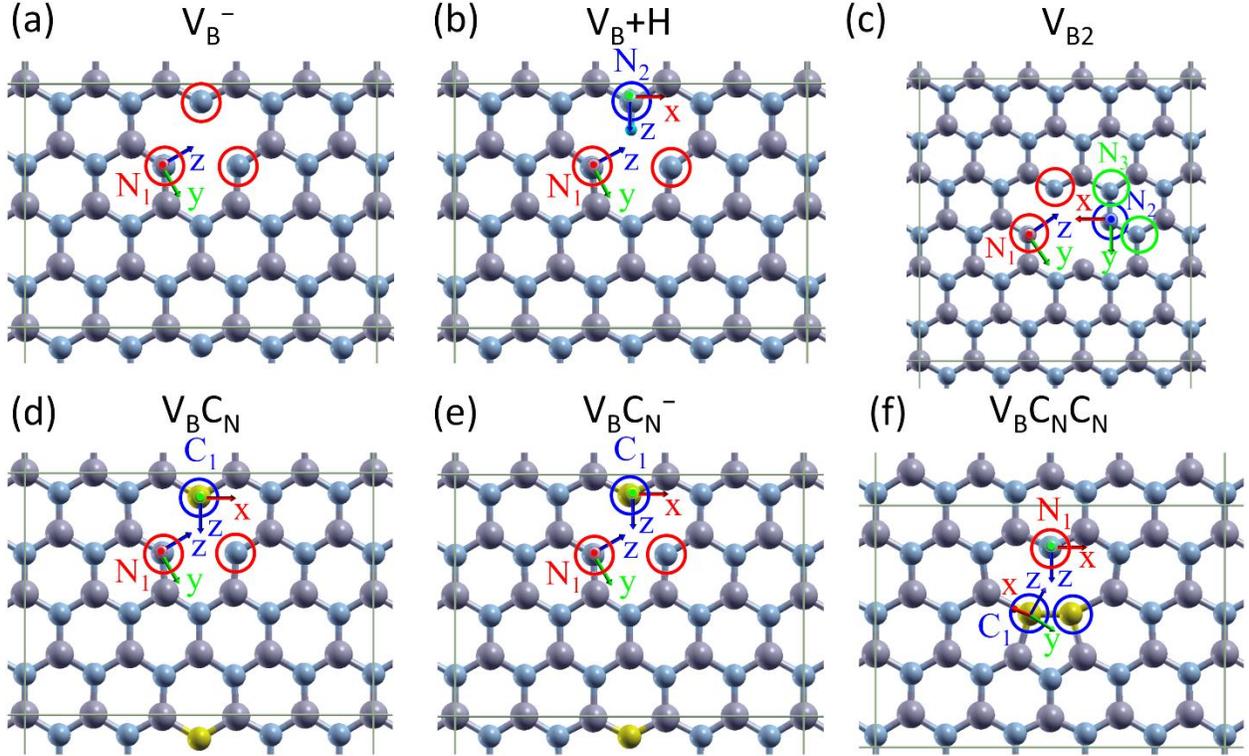

**Figure 1.** Periodic supercells of hBN with various $V_B$-based defect centers used in the GW-BSE calculations. (a) $V_B^-$, (b) $V_B$+H, (c) $V_{B2}$, (d) $V_BC_N$, (e) $V_BC_N^-$, (f) $V_BC_NC_N$. Note that in $V_BC_NC_N$ the nitrogen atom $N_1$ is tilted out-of-plane by 0.67 Å (see also Fig. 6a). Boron, nitrogen, hydrogen and carbon atoms are denoted by gray, dull blue, bright blue and yellow balls, respectively. The colored circles mark the atomic nuclei sites used for the calculation of the hyperfine coupling tensor A, with the same color denoting sites that are equivalent by symmetry. The principle axes of the hyperfine coupling tensor are marked at the atomic nuclei. In each case, $A_{zz}$ has the largest magnitude (see Table 2), and the z-direction points from the atomic nuclei site to the center of the defect except for the $N_2$ site of $V_{B2}$. Right-handed coordinate systems are used.

Table 1 shows the energies of the low energy bright excitons computed with GW-BSE for the defects in Figure 1. The energies of these excitons are much smaller than the optical onset in pristine ML hBN. As such, these excitons involve transitions between deep level defect states. Except for $V_B^-$, we find a number of these exciton energies well within the near infrared to visible spectral range, although their exact peak energies, polarization direction and spectral intensities differ (see below). These findings point to the potential of these defects as useful light emitters for applications. Excitation energies computed using the DMRG approach are also shown in Table 1. Although the cluster models used for the DMRG calculations are different from the periodic supercells used for the GW-BSE calculations, our primary focus is to search for additional low-energy excitations that are missing from the GW-BSE calculations. A more complete listing of the bright exciton energies and the DMRG excitation energies is provided in Table S1.

Table 1. Results from GW-BSE and DMRG calculations for boron vacancy-derived defects in ML hBN. For the GW-BSE calculations, we list the excitons with largest oscillator strength that contribute to the major optical absorption peaks. We focus on the low-energy spectrum up to the visible spectral range for all defects, except for $V_B^-$, where the lowest-energy bright excitons are listed. For the DMRG calculations, we focus on the lowest excitation energies. In parenthesis after the structure names, we denote their ground state spin multiplicity with T for triplets and D for doublets. See Table S1 for more results over a larger energy range.

| Defect type | BSE exciton energies (eV) | DMRG excitation energies (eV) |
| --- | --- | --- |
| $V_B^-$ (T) | 3.89, 3.92, 3.96 | 2.17, 2.18 |
| $V_B$+H (T) | 0.70, 1.24, 1.61, 1.87, 2.10, 2.49 | 1.01, 1.11 |
| $V_{B2}$ (T) | 2.86, 3.20 | 1.37 |
| $V_B C_N$ (T) | 1.22, 1.35, 1.78, 2.00, 2.18, 2.22, 2.56, 2.77 | 2.24, 2.80 |
| $V_B C_N^-$ (D) | 2.42 | 1.95, 1.98 |
| $V_B C_N C_N$ (D) | 1.43, 2.21, 2.35, 2.71, 2.78 | 1.05 |

The DFT results for the MR parameters are summarized in Table 2. Specifically, we report the hyperfine coupling (HFC) A-tensor, EPR *g*-tensor, and, for triplet states, the ground state zero field splitting (ZFS) parameters. For the HFC tensor calculation, we choose atomic nuclei sites close to the defect center, as indicated in Figure 1. Our calculations show that several of the atomic nuclei sites result in very similar HFC parameters. The equivalence of these sites is indicated in Figure 1 by colored circles around the sites. We note that atomic sites that are equivalent by structural symmetry also give very similar HFC parameters, as expected. However, similar HFC A-tensors are also obtained in sites that are not exactly equivalent by structural symmetry, such as for the indicated N sites in $V_{B2}$ (Figure 1c), which have very similar local environments. This information is helpful for comparison with angle-dependent EPR measurements.[14] For the HFC A-tensor calculations, we consider the $^{14}$N and $^{13}$C isotopes. The former is the prevalent naturally occurring isotope for N (99.6 %), while the latter is less common (1.1 %). Hyperfine coupling lifts the degeneracy of the electronic state. If the carbon atoms are spinless $^{12}$C isotopes, we expect the hyperfine coupling from the prevalent $^{14}$N isotopes to result in seven sublevels for $V_B^-$, five sublevels for $V_B C_N$ and $V_B C_N^-$ and three sublevels for $V_B C_N C_N$, since there are, respectively, three, two and one equivalent N atoms in each of these defect groups. For $V_B$+H, the HFC tensor elements are much smaller for $N_2$ than for $N_1$, so that the effect of HFC from $N_2$ will only be a perturbation to five sublevels due to HFC from the two $N_1$ sites. The size of the HFC tensor elements is characteristic of the defect.

In contrast to the HFC A-tensor elements, the variation in the *g*-tensor elements across different defect types is quite small (Table 2). ZFS refers to the energy splitting between the spin-sublevels in the absence of an external magnetic field. The ZFS parameters arise primarily from magnetic dipole-dipole interactions when there is more than one unpaired electron. Calculations for ZFS parameters are particularly useful for comparison with ODMR measurements on triplet ground state centers. Our predicted ZFS D parameter for $V_B^-$ is consistent with existing experimental measurements.[14]

We also compute the MR parameters for the neutral boron vacancy defect, $V_B$, which is not optically active under 3.1 eV, according to previous GW-BSE[34] and HSE06[50] calculations. This provides additional information to help distinguish defects, especially in ensemble measurements (e.g. EPR). We find that $V_B$ has similar ZFS parameters as $V_{B2}$, but $V_B$ and $V_{B2}$ can be distinguished by the distinct HFC parameters as well as their different in-plane symmetry ($D_{3h}$ for $V_B$ versus no symmetry for $V_{B2}$). For example, the $N_2$ site in $V_{B2}$ has substantially smaller hyperfine coupling parameters (see Table 2), while those for the $N_3$ sites are negligible.

Table 2. MR parameters for the boron vacancy-derived defects in ML hBN. The hyperfine coupling A-tensor nuclei sites are labeled in Figure 1, and $^{14}N$ and $^{13}C$ isotopes were considered in the calculation. The principle axes for A-tensor are also marked in Figure 1, while the principle axes for $g$-tensor are marked in Figure S1. ZFS parameters are not applicable for defects with fewer than two unpaired spins. The HFC tensor elements for the sites labeled by $N_3$ in the $V_{B2}$ defect are much smaller in magnitude than those for $N_1$ and $N_2$, and are omitted in this table.

| Defect type | Hyperfine coupling (HFC) tensor (MHz) | | | | g-tensor | | | Zero field splitting | |
|---|---|---|---|---|---|---|---|---|---|
| | Site | $A_{xx}$ | $A_{yy}$ | $A_{zz}$ | $g_{xx}$ | $g_{yy}$ | $g_{zz}$ | D/h(GHz) | E/h(MHz) |
| $V_B^-$ | $N_1$ | 47.5 | 46.0 | 88.1 | 2.0031 | 2.0031 | 2.0042 | 3.3 | 0.8 |
| $V_B$+H | $N_1$ | 34.6 | 11.1 | 44.6 | 2.0025 | 2.0030 | 2.0039 | 8.4 | 1084 |
| | $N_2$ | -3.8 | -4.2 | -4.5 | | | | | |
| $V_{B2}$ | $N_1$ | 48.3 | 46.9 | 75.0 | 2.0031 | 2.0036 | 2.0031 | -1.1 | -76.8 |
| | $N_2$ | 4.1 | 3.7 | 53.9 | | | | | |
| $V_B C_N$ | $N_1$ | 134.2 | 127.7 | 259.4 | 2.0031 | 2.0040 | 2.0030 | -1.4 | -412 |
| | $C_1$ | -26.7 | 69.2 | -60.8 | | | | | |
| $V_B C_N^-$ | $N_1$ | 299.8 | 288.9 | 547.7 | 2.0025 | 2.0043 | 2.0040 | n.a. | n.a. |
| | $C_1$ | -2.2 | -10.8 | -48.6 | | | | | |
| $V_B C_N C_N$ | $N_1$ | 90.7 | 100.2 | 481.3 | 2.0049 | 2.0023 | 2.0042 | n.a. | n.a. |
| | $C_1$ | 15.4 | 17.5 | 30.5 | | | | | |
| $V_B$ | $N_1$ | 37.5 | 26.4 | 54.7 | 2.0032 | 2.0031 | 2.0036 | -1.3 | -82.9 |

In the following, we discuss the electronic structure, optical properties and MR parameters of each of the boron vacancy-derived defect color centers (Figure 1) in detail.

1. Negatively charged boron vacancy $V_B^-$

The $V_B^-$ defect is a charged intrinsic single boron vacancy defect with $D_{3h}$ symmetry. There has been significant interest in the $V_B^-$ defect due to its intriguing electronic structure and magnetic response.[14, 24, 51] The $V_B^-$ defect is particularly complex from a theoretical point of view, as discussed in recent literature.[24, 37, 39] Our GW calculations predict several in-gap localized states for $V_B^-$ (see Fig. 2a). For the majority spin, they are occupied and located ~1.0 eV above the hBN valence band maximum (VBM). For the minority spin, four states close to the VBM are occupied while two unoccupied deep-level defect states are located ~1.0 eV below the conduction band minimum (CBM). The orbital characters of these defect states are labeled in Fig. 2a (for detailed orbital wavefunction plots, see Fig. S3). In Fig. S4, we show the defect levels obtained for the same system, using a DFT calculation with the hybrid HSE exchange-correlation functional,[52] an approach commonly used in the literature.[23, 24] We see that the HSE results are similar to the GW predictions for the majority spin states, but the defect level positions and orbital ordering are different for the minority spin states. This highlights the importance of dynamical non-local correlation effects present in the GW calculations.

The optical absorption spectrum for $V_B^-$, as predicted by GW-BSE calculations, is shown in Fig. 2b. The onset of optical transitions starts at ~3.9 eV, and arises from transitions between the HOMO-3 ($a_1'$) and LUMO/LUMO+1 ($e'$) quasiparticle levels (HOMO: highest occupied molecular orbital; LUMO: lowest unoccupied molecular orbital) in the spin minority component. The corresponding optical absorption spectra for pristine monolayer hBN is also plotted in Figure 2b. In order to predict additional low-energy optical transition states with multi-reference character,[24, 44] we compute the vertical excitation energies of the $V_B^-$ defect using the DMRG method (see Table 1). We found two lower energy, almost degenerate excitations at 2.17 and 2.18 eV, consistent with the converged results of a recent detailed study on DMRG calculations for $V_B^-$.[44]

For the MR parameters of $V_B^-$, we obtain a three-center $^{14}$N HFC tensor A of [47.5, 46.0, 88.1] MHz. The radial isotropic component at ~47 MHz is in good agreement with the experimental result from Ref. 14. The isotropic in-plane *g*-tensor value is 2.0031 (see Table 2). We obtain the zero field splitting (ZFS) D parameter equal to 3.3 GHz, and E is close to zero within numerical accuracy (see Table 2). The D value is in good agreement with the literature values.[14, 24, 37] An anisotropic E value at 50 MHz was obtained in experiment,[14] and we find that a uniaxial tensile strain of about ~3.6% can account for this anisotropy (see Fig. S4). The close agreement between our computed MR parameters and known experimental measurements as well as previous theoretical results (see Table S2) validates our first principles approach, and corroborates our predictions on defect species that have not been experimentally identified.

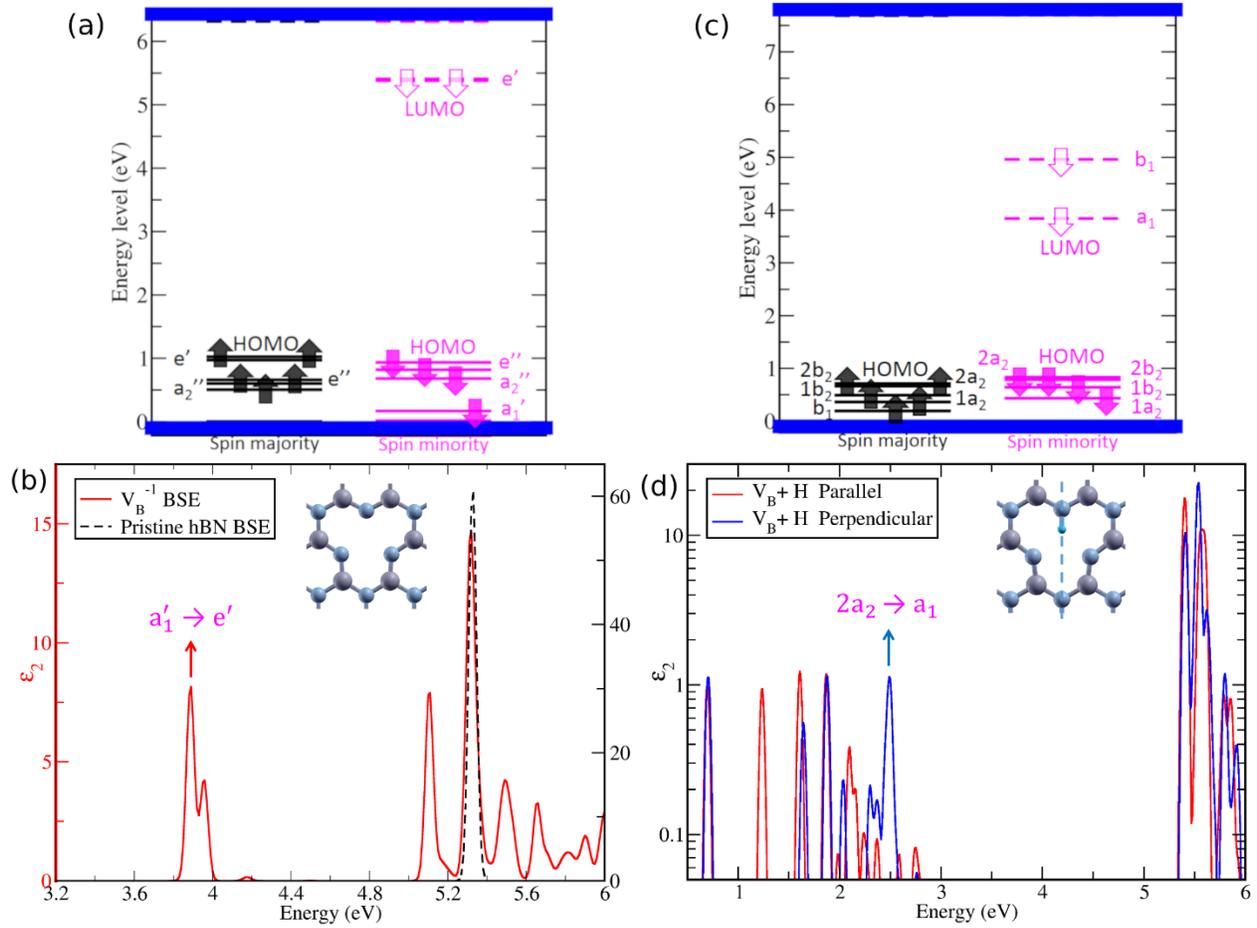

Figure 2. (a) Ground state GW level diagram of the $V_B^-$ triplet defect at Γ point, and (b) the corresponding GW-BSE optical absorption spectrum. (c) and (d) are analogous to (a) and (b) but for the $V_B$+H triplet defect. Note that the vertical axis of (d) is in logarithmic scale. In (b) we also plot the pristine ML hBN spectrum for comparison. In the level diagram plots, filled arrows denote occupied states, while empty arrows denote unoccupied states. The VBM is set to zero on the energy axis, and the blue colored regions denote the onset of bulk bands. In the insets of the spectra plots are geometric models of the respective defect centers. These descriptions apply to all level diagrams and atomic structures presented in this work. In (d), 'parallel' and 'perpendicular' refer to the absorption of light polarized parallel and perpendicular to the symmetry axis labeled in the inset by the dashed line.

2. Hydrogen-passivated boron vacancy $V_B$+H

It is possible that the single boron vacancy has one or several of the surrounding nitrogen dangling bonds passivated by hydrogen. In particular, passivation of the boron vacancy by a single hydrogen atom gives rise to the $V_B$+H defect. The $V_B$+H defect has an explicit in-plane anisotropy, reducing the local defect symmetry into $C_{2v}$, with a symmetry axis passing through the boron vacancy and hydrogen sites. The GW quasiparticle levels for this defect are similar to $V_B^-$ (see Fig. 2c), but with several differences. Besides the expected changes in symmetry labels of the defect states, the two unoccupied spin down states are now deeper (closer to mid-gap) and split, being at 2.7 and 3.8 eV lower than the CBM. The occupied mid-gap states are still close to the VBM but with slightly different energies.

Due to these altered in-gap state positions and optical transition selection rules, the GW-BSE calculations show that there are several weaker optical transitions within the lower energy range for the $V_B$+H defect. The spectral positions of these peaks are around 0.70, 1.24, 1.61, 1.87, 2.10 and 2.49 eV (see Table 1 and Fig. 2d). These numbers are corroborated by DMRG excitation energies (see Table 1 and S1), indicating a rich number of optical excitations. The optical absorption spectra is also polarization-dependent as shown in Fig. 2d.

The $V_B$+H defect may be responsible for several SPE centers already observed in hBN systems. In particular, the ~2.49 eV peak which stems from HOMO ($2a_2$) to LUMO ($a_1$) transition (see Fig. 2d and Fig. S5) in the spin minority channel, is a potential candidate for the widely observed ~2 eV SPE line.[9, 12, 13, 25, 26, 46] In particular, emission at 2.1 eV was observed in a recent experiment where polarization-dependent measurements give rise to intensity maxima and minima for polarizations that are perpendicular to one another, with the intensity minima reaching nearly zero values.[29] These observations are consistent with our prediction that the oscillator strength for this transition is nearly zero for light polarized parallel to the symmetry axis but is significant for light polarized perpendicular to the axis. This defect is potentially of great interest due to the observation of a giant in-plane Stark effect in these same experiments.[29] Our assignment is further corroborated by the polarization-dependent emission results for the first detection of quantum emission from hBN, corresponding to SPE at 2.0 eV.[9] While this SPE at 2.0 eV was originally assigned to the defect complex, $V_NN_B$, the formation energy of $V_NN_B$ (> 8 eV) is much larger than that of $V_B$+H (< ~ 5 eV). Given that SPE at 2.0 eV has been observed in several experimental works,[9, 12, 13, 25, 26] this emission is unlikely to arise from a defect with an exceptionally large formation energy.

We note that several lower energy peaks in the optical absorption spectra are also related to the HOMO ($2a_2$) to LUMO ($a_1$) transition. Our predictions suggest other energy ranges for the potential detection of SPE from the $V_B$+H defect, including the spectrally-isolated peak at ~1.24 eV which is close to the telecom O-band. The peak at ~1.24 eV also exhibits polarization dependence, but in contrast to the peak at ~2.49 eV, its oscillator strength is largest for light polarized parallel to the polarization axis.

For the $V_B$+H defect, we predict a HFC A of [34.6, 11.1, 44.6] MHz on the two surrounding dangling bond nitrogen sites. Hydrogen passivation of the remaining $^{14}$N nucleus at the defect site reduces the values of A to [-3.8, -4.2, -4.5] MHz on the passivated N atom, indicating negligible contribution to the overall hyperfine splitting. This is distinct from the three-center $V_B^-$ case and will likely result in a different number of hyperfine sublevels (~five versus seven for $V_B$+H and $V_B^-$, respectively). The anisotropy of the in-plane $g$-tensor (see Table 2) and relatively large ZFS E parameter of 1084 MHz also reflect the in-plane anisotropy of the defect. The ZFS D parameter is 8.4 GHz, more than twice the value for $V_B^-$. Since hydrogen atoms are generally difficult to identify in experiments, these differences in the MR parameters are important in distinguishing between the $V_B$+H and $V_B^-$ defects. Furthermore, they are also different from the MR parameters of the neutral $V_B$ defect (see Table 2). Using controlled experiments involving carbon implantation, recent studies have shown that carbon impurities also result in emission at ~2eV, which has been attributed to the $V_BC_N^-$ defect.[27] The $V_B$+H defect can be distinguished from the $V_BC_N^-$ defect by its large ZFS parameters, compared to $V_BC_N^-$ which has a doublet ground state with zero ZFS parameters. In our discussion below, we also suggest that the $V_BC_NC_N$ defect can also be responsible for a ~2 eV emission in carbon-enriched samples.[28]

3. Double boron vacancy $V_{B2}$

The $V_{B2}$ defect consists of two neighboring boron vacancy sites located next to a nitrogen/boron anti-site (Fig. 1c). It could be generated by additional boron loss or higher-dose beam irradiation. Alternatively, it could be viewed as a merging of $V_B$ and $V_NN_B$ defects, with the latter being widely studied in literature.[9,53] The antisite N-induced N-N bonds are found to greatly stabilize this divacancy.[54] There is now no local symmetry around the defect center. GW calculations predict occupied and unoccupied in-gap states from $V_{B2}$ for both spin components (Fig. 3a). Compared with $V_B$+H and $V_B^-$, the occupied defect levels for $V_{B2}$ are higher in energy with respect to the VBM. Fig. S6 shows the orbital character of the defect states.

From BSE calculations, we predict lower energy exciton transitions at 2.86 and 3.20 eV, and a series of exciton energies above 3.5 eV (see Fig. 3b and Table 1). They are corroborated by the DMRG-predicted excitation energies at 3.04, 3.75 eV and higher. The 2.86 eV peak stems from the HOMO-1 to LUMO transition in the spin majority channel, while the 3.20 eV peak arises from the HOMO to LUMO+1 transition in the spin minority channel. Both of these peaks are well-separated in energy from other states, indicating their potential as SPEs in the visible range. The 2.86 eV exciton in $V_{B2}$ is a likely candidate for the recently observed SPEs in hBN under substantial electron beam irradiation, where several ZPLs at energies ~2.80 eV are observed.[30] Furthermore, our polarization dependent transition amplitudes in Fig. 3c indicate a clear linear polarization, agreeing with the experimentally observed linear polarization from the same work.[30] With 0° along the x-axis in Fig. 3b (a lattice vector in pristine hBN), we see that the dipole moment is nearly zero for light polarized along 90°. The dependence is asymmetric about 90°, similar to the experimental observation[30] that the polarization angle of emission does not coincide with a crystal axis. The polarization dependence for the 3.20 eV exciton is opposite to that of the 2.86 eV exciton (Fig. 3c).

The HFC tensor elements for the two nitrogen sites $N_1$ closest to the defect center (Fig. 1) are of similar magnitude as those for $V_B$+H and $V_B^-$. The HFC $A_{xx}$ and $A_{yy}$ tensor elements for $N_2$ (farther from the defect center) are rather small in magnitude (~4 MHz), but the $A_{zz}$ term is ~54 MHz (see Table 2). Like $V_B$+H, the anisotropy of the defect geometry can be seen in the in-plane anisotropy of the *g*-tensor and the large magnitude of the ZFS E parameter. Compared to $V_B$+H, we predict that $V_{B2}$ has a ZFS D parameter that has a negative sign and smaller magnitude (see Table 2). These characteristics could serve as fingerprinting labels to experimentally identify the $V_{B2}$ defect. The ZFS parameters for $V_{B2}$ are quite close to those of the neutral $V_B$ defect, but the HFC tensor values and symmetry are distinct between these two cases.

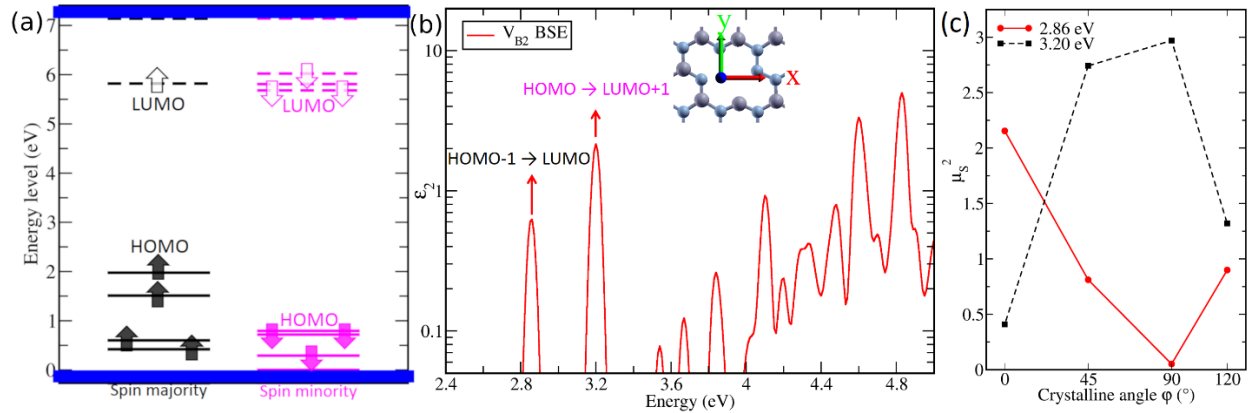

Figure 3. (a) Ground state GW level diagram for the $V_{B2}$ triplet defect, and (b) the corresponding GW-BSE optical absorption spectrum with polarization along 45° direction, where 0° direction is referenced horizontally along the x-axis as indicated in the inset. (c) Polarization dependent $\mu_S^2$, the square of the optical dipole matrix element, for the 2.86 and 3.20 eV exciton modes.

4. Boron vacancy carbon/nitrogen substitution $V_BC_N$

Extrinsic impurity elements such as carbon can be incorporated in the hBN lattices during the synthesis or processing steps.[27, 28, 53, 55] The simplest boron vacancy-related extrinsic defect with carbon is a double defect complex with one neighboring nitrogen site substituted by carbon, i.e. $V_BC_N$. The local symmetry is $C_{2v}$, similar to the $V_B$+H case, with the symmetry axis passing through the vacancy and the carbon sites. The computed GW energy level diagram for this defect is shown in Fig. 4a. There are several occupied and unoccupied single particle levels within the band gap for both spin components. The unoccupied defect levels are deep in the gap (at least 2 eV from the CBM). The GW-BSE optical spectra has peaks at 1.22, 1.35, 1.78 and 2.00 eV, as well as a number of peaks between 2.2 and 3.5 eV, with distinct polarization directions (see Fig. 4b). The peak with largest oscillator strength is at ~1.78 eV, and arises from the HOMO-1 ($b_1$) to LUMO ($a_2$) transition in the spin minority channel. Another peak with large oscillator strength is at ~2.00 eV. The peak at ~2.00 eV arises from a combination of transitions: the HOMO-2 ($b_1$) to LUMO ($a_1$) transition in the spin majority channel, and the HOMO ($1b_2$) to LUMO ($a_2$) and HOMO ($1b_2$) to LUMO+1 ($2b_2$) transitions in the spin minority component. Figure S7 shows the orbital characters of the defect states.

For the magnetic resonance parameters of $V_BC_N$ defect, we predict large hyperfine tensor elements associated with naturally occurring $^{14}$N (A = [134.2, 127.7, 259.4] MHz) while the hyperfine tensor elements for $^{13}$C are smaller in magnitude (A = [-26.7, 69.2, -60.8] MHz). There are two N atoms at the defect center. The $V_BC_N$ defect has a triplet ground state, and the ZFS D parameter is -1.4 GHz while the ZFS E parameter is -412 MHz, consistent with the in-plane anisotropy of the defect. The $g$-tensor also possesses in-plane anisotropy (see Table 2). The extrinsic character of the $V_BC_N$ defect and its promising spectroscopic properties point to the importance of this defect, which could be amenable to intentional creation and manipulation in the future.

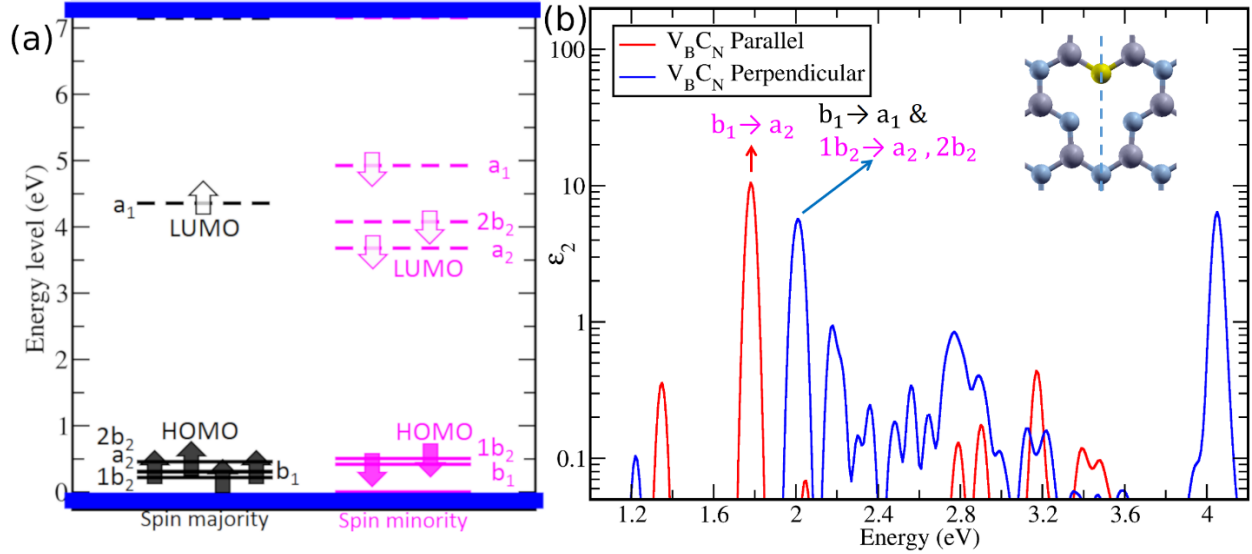

Figure 4. (a) Ground state GW level diagram for the $V_BC_N$ triplet defect, and (b) the corresponding GW-BSE optical absorption spectrum. In the inset of the spectral plot is geometric model of the defect center, with the polarization reference axis indicated with a dashed line.

5. Negatively charged boron vacancy carbon/nitrogen substitution $V_BC_N^-$

Recent work suggests that the negatively charged defect $V_BC_N^-$ could be important for interpreting spectroscopic properties of SPEs in hBN.[27] Compared with the charge neutral case, one extra spin minority state is occupied by the extra charge (see Fig. 5a). The GW calculations also predict that the occupied defect levels have higher energies with respect to the VBM, compared to the neutral $V_BC_N$ defect, due to on-site charging effects. In contrast to the neutral $V_BC_N$ defect, the GW-BSE spectrum for $V_BC_N^-$ has only a few peaks in the energy range shown (see Fig. 5b). There is a spectrally-isolated bright exciton state at 2.42 eV for light polarized perpendicular to the symmetry axis. This is the only peak within the visible spectrum, and stems from the HOMO-1 ($b_1$) to LUMO ($a_1$) transition (see Fig. S8) in the spin majority component.

For the MR parameters, we predict that the hyperfine tensor elements associated with naturally occurring [14]N are larger than those for the neutral $V_BC_N$ defect (A = [299.8, 288.9, 547.7] MHz) while the hyperfine tensor elements for [13]C are smaller in magnitude (A = [-2.2, -10.8, -48.6] MHz). The g-tensor displays in-plane anisotropy (see Table 2). In addition, since the $V_BC_N^-$ defect is a spin doublet state, there are no sizable ZFS parameters, which can be crucial in differentiating between the negatively charged and charge neutral cases. Alternatively, the charge state of $V_BC_N$ defect could be tuned by gating or charge transfer. We note that the zero ZFS parameters and the ~2.42 eV GW-BSE optical absorption peak (together with the ~2.55 eV excited state energy predicted by DMRG) corroborate well with the assignment of the ~2 eV emission line to $V_BC_N^-$ in Ref. 27, where hBN was prepared with carbon implantation and no sizable ZFS was measured. We note that the $V_BC_NC_N$ doublet defect is also a possible candidate consistent with these observations (see below). The $V_BC_N^-$ and $V_BC_NC_N$ defects can be distinguished from the fact that $V_BC_N^-$ has two N atoms at the defect center, thus resulting in five sublevels from hyperfine splitting, while $V_BC_NC_N$ has only one N atom at the defect center, resulting in three sublevels from hyperfine splitting.

6. Boron vacancy double carbon/nitrogen substitution $V_BC_NC_N$

With a higher carbon concentration, additional $C_N$ sites around the defect center are likely. The $V_BC_NC_N$ defect consists of a boron vacancy surrounded by two $C_N$ sites (see Fig. 1). The two carbon impurities actually form a 5-member ring with their surrounding atoms, with a C-C bond stabilizing the defect. The C-C bond length is 1.48 Å, which is slightly larger than the graphene C-C bond length of 1.42 Å, but smaller than the typical bond length for a C-C single bond (~1.54 Å). Our Bader charge analysis[56] indicates that each of the C atoms, as well as $N_1$ (see Fig. 1f and 6a), gains about one electron from each of their boron nearest neighbors. It is observed that $N_1$ is tilted out-of-plane by 0.67 Å in the ground state geometry (Fig. 6a). Based on the Bader charge analysis, we suggest that this out-of-plane distortion is not due to additional charge transfer to $N_1$ but is likely arising from steric hindrance. This tilting results in a $C_S$ local symmetry, with the image plane perpendicular to the material plane and passing through the boron vacancy and $N_1$ sites, suggesting that the emission energy of this defect can be controlled with an out-of-plane electric field.[57] The orbital characters of the in-gap states are shown in Fig. 6b-c, where C-C bonding and antibonding features can be observed.

The GW energy level diagram for this defect is shown in Fig. 5c. Compared to the other defects considered in this work, the number of mid-gap states is smaller in the $V_BC_NC_N$ defect due to the smaller number of valence electrons in carbon compared to nitrogen and the formation of the C-C bond. Our BSE calculations predict exciton transitions at 1.43, 2.21, 2.35, 2.71, 2.78 eV and higher, with distinct optical polarization directions (see Fig. 5d). The 1.43 and 2.21 eV peaks are due to the HOMO (2a') to LUMO (3a') transition in the spin minority component, while the 2.35 eV peak stems from the HOMO (3a') to LUMO (a'') transition in the spin majority component. We note that the DMRG calculations also predict another lower energy excited state at 1.05 eV, which is far from any excitonic energy in the GW-BSE calculations, thus suggesting an additional excitation of multi-reference character. Depending on the brightness of optical transition corresponding to this excited state, $V_BC_NC_N$ may be useful for applications requiring telecom wavelengths.

In recent experiments on carbon-enriched hBN samples,[28] a 1.995 eV zero-phonon line was observed, with the photoluminescence excitation (PLE) spectrum for this emission showing a contribution from excitations at ~2.5 eV with a shoulder at lower energies and another peak at ~2.2 eV. Based on the GW-BSE optical absorption spectra for the carbon-based defects considered here, we suggest that $V_BC_NC_N$ is a likely candidate defect responsible for the observed emission. In particular, there are optical absorption peaks at ~2.2 eV and ~2.4 eV for $V_BC_NC_N$ (Figure 5d). In the same experiment, a 1.54 eV emission doublet was also observed. The corresponding PLE spectra have excitations at ~2.4 eV and ~1.7 eV. As $V_BC_N^-$ and $V_BC_N$ have optical absorption peaks at 2.4 eV (Figure 5b) and 1.8 eV (Figure 4b), respectively, they are possible candidates for the 1.54 eV emission doublet.

As discussed above, the hyperfine splittings can be used to distinguish the $V_BC_NC_N$ and $V_BC_N^-$ defects. Besides the distinct number of hyperfine splitting sublevels, we also find that the in-plane HFC tensor elements are significantly smaller in $V_BC_NC_N$ compared to those in $V_BC_N^-$ (see Table 2). The *g*-tensor is similar to the $V_BC_N^-$ case, exhibiting in-plane anisotropy. The ZFS parameters are zero for $V_BC_NC_N$ as it is a spin doublet defect.

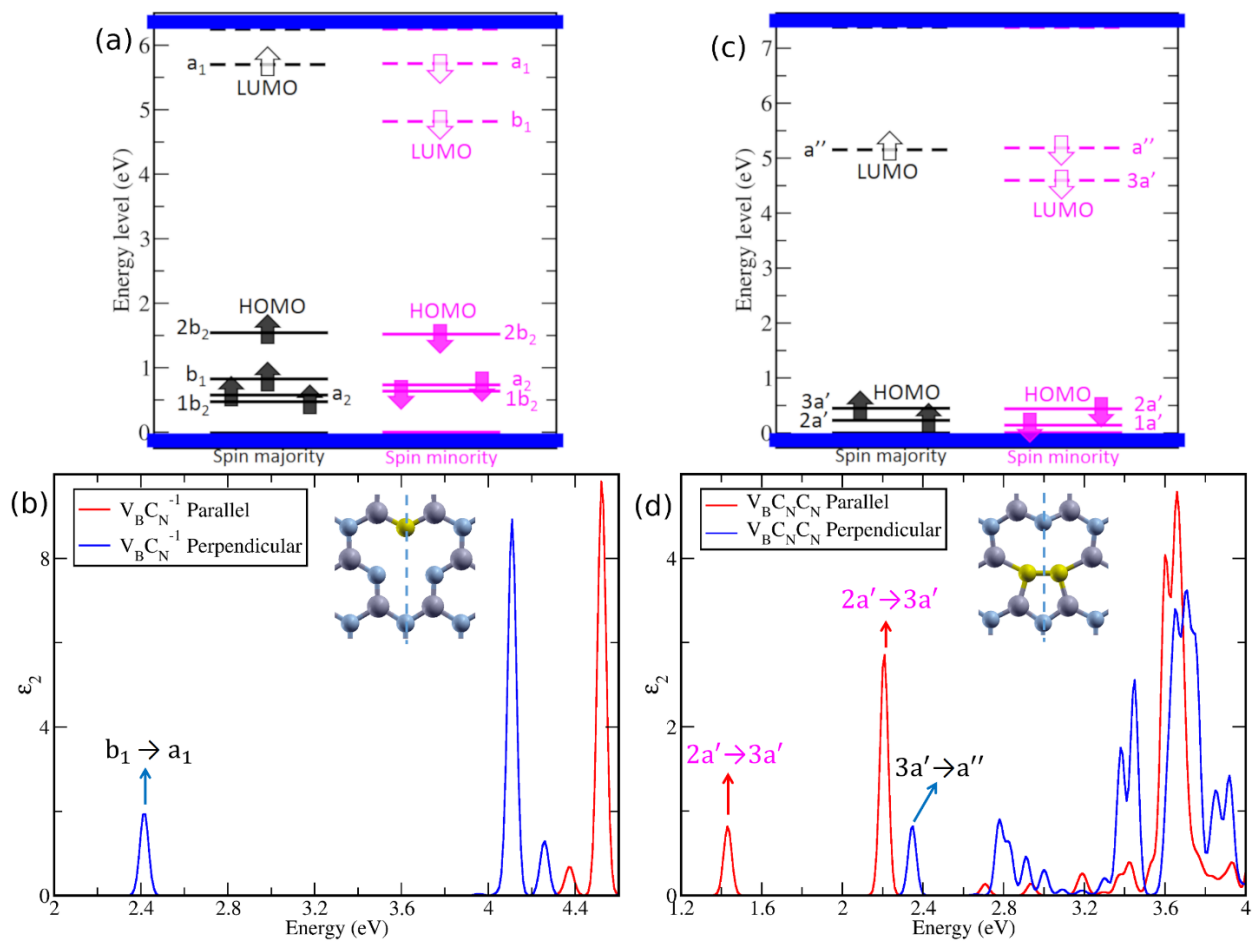

Figure 5. (a) Ground state GW level diagram of the $V_BC_N^-$ doublet defect, and (b) the corresponding GW-BSE optical absorption spectrum. (c) and (d) are analogous to (a) and (b) but for the $V_BC_NC_N$ doublet defect. In the insets of (b) and (d) are geometric models of the respective defect centers, with the polarization reference axes indicated by dashed lines.

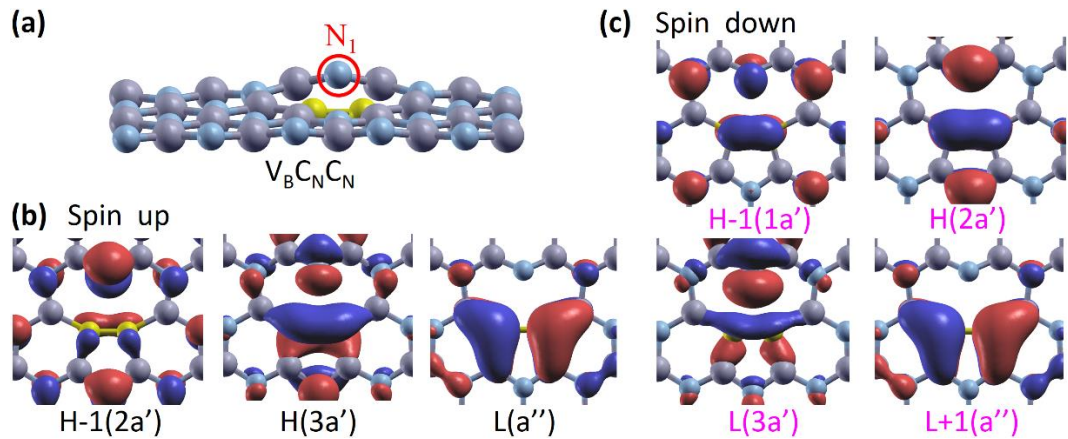

Figure 6. (a) Atomic structure of $V_BC_NC_N$, showing the nitrogen atom $N_1$ tilting out of plane, and (b) spin up and (c) spin down electronic orbital wavefunction plots for in-gap levels. 'H' stands for HOMO and 'L' for LUMO.

**Conclusion**

Defect centers in hBN are promising in overcoming several critical limitations of SPEs in three dimensional systems.[2, 4] However, progress in experimental synthesis, especially with regards to controllable defect creation and manipulation in hBN, is still at an early stage. While defects can be intentionally created in hBN using various approaches, a detailed knowledge of the photophysical characteristics of these defects is still lacking. The ability to identify the underlying atomic structure is critical to advancing our understanding of how to control and manipulate defects, as has been the case of the NV$^-$ center in diamond. These considerations underscore the important role of first principles calculations in uncovering the structure-property relationships in these defects. Of particular importance are accurate and predictive calculations of quantities that can be measured experimentally in order to identify the defects through spectroscopy.

In this work, we predict, using state-of-the-art first principles calculations, the photo-physical properties of six boron vacancy ($V_B$)-derived localized defect emitting centers in monolayer hBN. Based on the optical spectra alone, we identify several potential candidates for SPE at ~ 2 eV. These include the $V_B$+H spin triplet as well as $V_BC_N^-$ and $V_BC_NC_N$ spin doublet defects. All of these defects exhibit anisotropic polarization dependence. Although the different experiments report emission from defects in hBN at approximately the same energy (~2 eV), the experimental conditions and sample preparation procedures are different across experiments.[9, 12, 13, 25-29] Some experiments specifically perform carbon implantation[27] or incorporate carbon impurities[28] during the growth process. Based on the sample preparation conditions, it is likely that the $V_B$+H defect is relevant to recent measurements of a giant polarization-dependent in-plane Stark effect of the ~2eV SPE.[29] Besides growth conditions, the different candidates can be distinguished by their magneto-resonance parameters. The $V_B$+H defect has distinctive ZFS parameters, with a large E value of 1084 MHz and a D value of 8.4 GHz, in contrast to the $V_BC_N^-$ and $V_BC_NC_N$ defects which have zero ZFS. Photoluminescence excitation spectra can also provide more direct comparison with the energies of the optical absorption peaks predicted in our calculations. Based on this comparison, we suggest that $V_BC_NC_N$ defect is a likely candidate for the ~2eV emission observed in recently synthesized carbon-enriched hBN,[28] while the observed 1.54 eV emission doublet from the same work is likely from $V_BC_N^-$ and $V_BC_N$ . Our calculations also shed light on recent experimental progress in creating position-controlled quantum emitters in electron-beam-irradiated hBN.[30] The observed emission at 2.8 eV is in excellent agreement with our observation of a sharp, spectrally isolated exciton at 2.86 eV in $V_{B2}$. This assignment is highly likely given that boron vacancies are preferentially created through electron-beam irradiation of hBN,[30] and the observed linear polarization of emission in both theory and experiment. While further confirmations with ODMR measurements are highly desirable, such an assignment of position-controllable SPEs in hBN paves the way to developing a more in-depth understanding of the SPEs so that their control and manipulation can be enhanced.

Besides providing predictions on useful spectroscopic and MR information to identify emitting centers in experiments, our calculations also serve to guide experiments to design defects from a bottom-up approach for specific applications. We predict the presence of spectrally-isolated lower-energy absorption peaks for $V_B$+H, $V_BC_N$ and $V_BC_NC_N$ defects, including one at ~1.2 eV for $V_B$+H, which renders them potentially useful for applications requiring telecom wavelengths. We find that $V_B^-$, $V_B$+H, $V_{B2}$ and

$V_BC_N$ are all defects with triplet ground states and non-zero ZFS parameters, which are promising for the optical initialization of the defect centers. The intrinsic exciton radiative decay rates of these defects from our GW-BSE calculations[58, 59] are plotted on the same scale in Figure 7. These radiative decay rates range from $1\times10^8$ to $1\times10^{10}$/s, showing high count rates, which renders them useful as localized emitters. Finally, the out-of-plane dipole moment in $V_BC_NC_N$ suggests that the emission from $V_BC_NC_N$ can be tuned by a gate voltage.

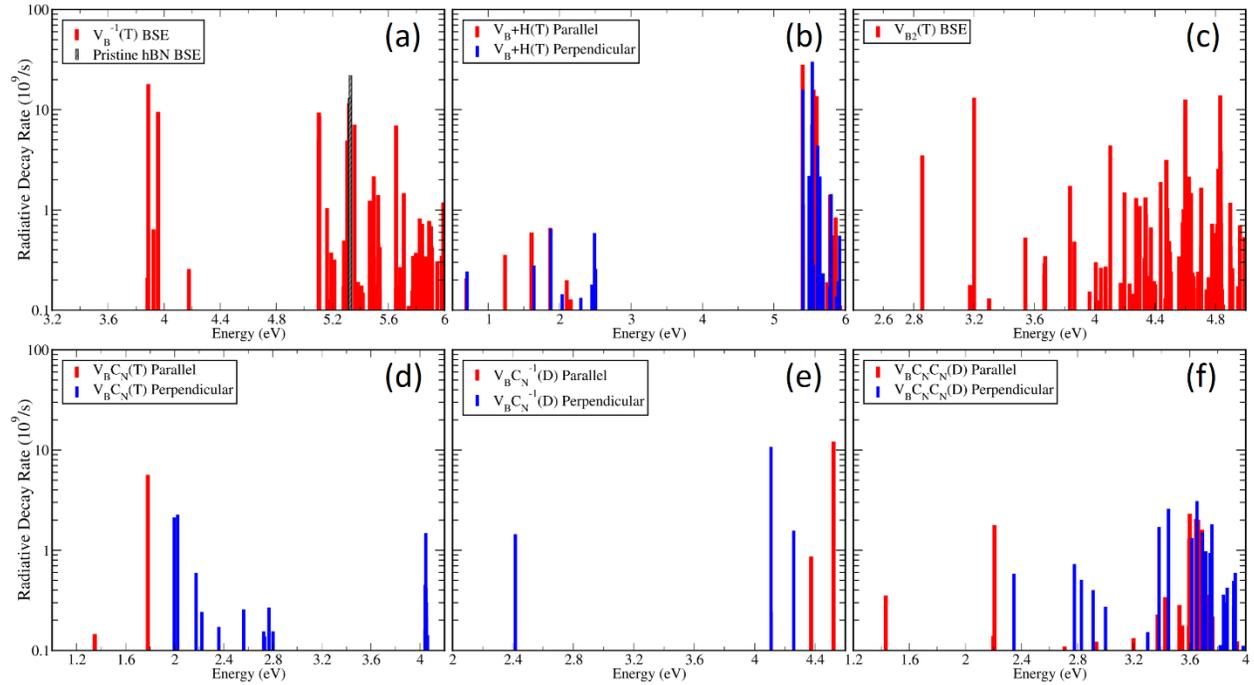

Figure 7. Intrinsic exciton radiative decay rates from the GW-BSE calculations for defect centers in hBN. (a) $V_B^-$, (b) $V_B+H$, (c) $V_{B2}$, (d) $V_BC_N$, (e) $V_BC_N^-$ and (f) $V_BC_NC_N$. In parenthesis, we denote triplets as T and doublets as D.

The number of defect centers studied here is small compared to the number of possibilities of defect centers within the hBN lattice.[12, 60] However, previous theoretical studies have shown that many of the intrinsic defects have very high formation energies.[46] The defects that we have focused on here have relatively lower formation energies. Another important consideration is the ability to control the synthesis of the defects in order to create localized emitters with specific and known characteristics. $V_B$-derived defects are examples of defects that can be created intentionally, because of the prevalence of boron vacancies in hBN samples treated with electron bombardment.[47] The photophysical characteristics of defects involving carbon are also relevant for advancing the objective of manipulating intentionally-created defects, as several experimental groups have incorporated carbon into the hBN lattice.[27, 28, 55] Future work in accurately determining the excited state relaxation and reorganization energies for the promising defects are highly desirable,[35, 61] since experimental works mostly report optical emission energies. However, it is known that the emission energies are slightly smaller than the absorption energies, by hundreds of meVs.[3, 20, 24, 37-39] The peak positions in the optical absorption spectra can also be directly compared with photoluminescence excitation spectra to guide the identification of the defect centers, as discussed above. Collectively, our results provide important information on the photophysical characteristics of boron-vacancy-derived defects in hBN. This information not only serves to identify the SPEs already observed in experiment, but also enables the bottom-up design of localized emitters in hBN for target applications.

**Methods**

First principles GW-BSE: We use the BerkeleyGW package[31] to carry out our first principles GW-BSE calculations, using mean-field DFT PBE[62] wavefunctions from the Quantum ESPRESSO package. We performed one-shot $G_0W_0$ calculations using the slab Coulomb truncation approach,[63] the Hybertsen-Louie generalized plasmon pole model,[64] and the static remainder method[65]. For the supercells involving defects (see Fig. 1), we used a 10 Ry dielectric matrix cutoff and 1200 bands for computing the dielectric matrix as well as the self-energy corrections. The non-uniform neck subsampling[66] technique was used in conjunction with a k-mesh of 4×4×1 for $V_{B2}$, and 6×9×1 for the rest. The BSE equation is solved by direct diagonalization within the Tamm-Dancoff approximation, including bands with energy at least 1 eV beyond the bulk band edges, and the BSE fine grid used was 3 times of the k-mesh used in the GW calculations. For the pristine ML hBN in the orthogonal 1×2 supercell, we used a 10 Ry dielectric cutoff and 1000 bands for computing the dielectric matrix as well as the self-energy corrections. The non-uniform neck subsampling[66] technique was used in conjunction with a 15×9×1 k-mesh for the GW calculations. A 75×45×1 fine k-mesh was used for the BSE calculations.

DMRG calculations: The DMRG calculations are done with the PySCF[42] set-up and the BLOCK code[41] on hydrogenated two-layer ring cluster models of hBN defect centers. The active spaces used are tabulated in the supplementary information. We use the unrestricted Hartree-Fock orbitals with ccpvdz basis sets, a state averaged ES calculation and a maximum subspace size M of 2000.

MR parameters: We carry out all-electron EPR and ZFS calculations by employing the ORCA code[43] on hydrogenated hBN clusters with three layers of rings around the defect centers. The input setup for hyperfine coupling and *g*-tensor is: ! PBE0 IGLO-II AUTOAUX. The input setup for the ZFS D and E parameter calculation describing spin dipole-spin dipole interaction is: ! UKS B3LYP def2-TZVP def2/J SOMF(1X).

Radiative decay rates: To compute exciton radiative decay rates for linearly polarized transitions, we use $\gamma_S(\mathbf{0}) = \{4\pi e^2 E_S(\mathbf{0})\mu_S^2\}/\{\hbar^2 c A_{sc}\}$ following Ref.[59], where $\mu_S^2 = \frac{\{\hbar^2|\langle G|p_{\hat{e}}|\Psi_S(\mathbf{0})\rangle|^2\}}{\{m^2 E_S^2(\mathbf{0})N_{\mathbf{k}}\}}$ is the square of the dipole matrix element, $m$ is the electron mass, $N_{\mathbf{k}}$ is the number of **k**-points, $S$ denotes the exciton mode, $E_S(\mathbf{0})$ is the exciton energy at zero momentum, and $A_{sc}$ is the supercell area. For non-polarized transitions, we use: $\gamma_S(\mathbf{0}) = \{8\pi e^2 E_S(\mathbf{0})\mu_S^2\}/\{\hbar^2 c A_{sc}\}$ following Ref.[58].

**Acknowledgements**

We acknowledge funding support from Ministry of Education, Singapore under grant R-144-000-421-112. We gratefully acknowledge computational resources at the Centre of Advanced 2D Materials (CA2DM) and at the National Supercomputing Centre (NSCC) in Singapore. In addition, the computational work was supported by CA2DM, funded by the National Research Foundation, Prime Minister's Office, Singapore, under its Medium-Sized Centre Programme.

**Supporting Information Available**: Defect center clusters, strain tuning of the $V_B^-$ ZFS E parameter, full table of BSE exciton, DMRG ES energies and DMRG active space size, MR parameters of $V_B^-$ in comparison with literature and electronic orbital wavefunction plots for the defect levels. This material is available free of charge via the Internet at http://pubs.acs.org

TOC Graphic

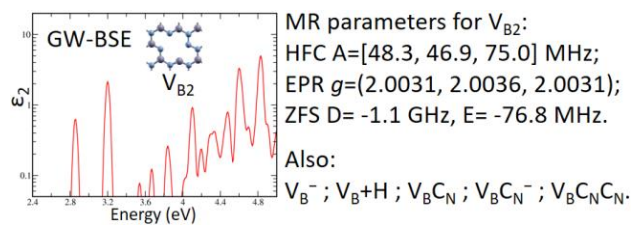

# Supplementary Information for:

# Photo-Physical Characteristics of Boron Vacancy-Derived Defect Centers in Hexagonal Boron Nitride

Yifeng Chen[1,2] and Su Ying Quek[1,2,3,4]

1. Department of Physics, National University of Singapore, 2 Science Drive 3, Singapore 117551

2. Centre for Advanced 2D Materials, National University of Singapore, 6 Science Drive 2, Singapore 117546

3. NUS Graduate School Integrative Sciences and Engineering Programme, National University of Singapore, Singapore 117456

4. Department of Materials Science, National University of Singapore, 9 Engineering Drive 1, Singapore 117575

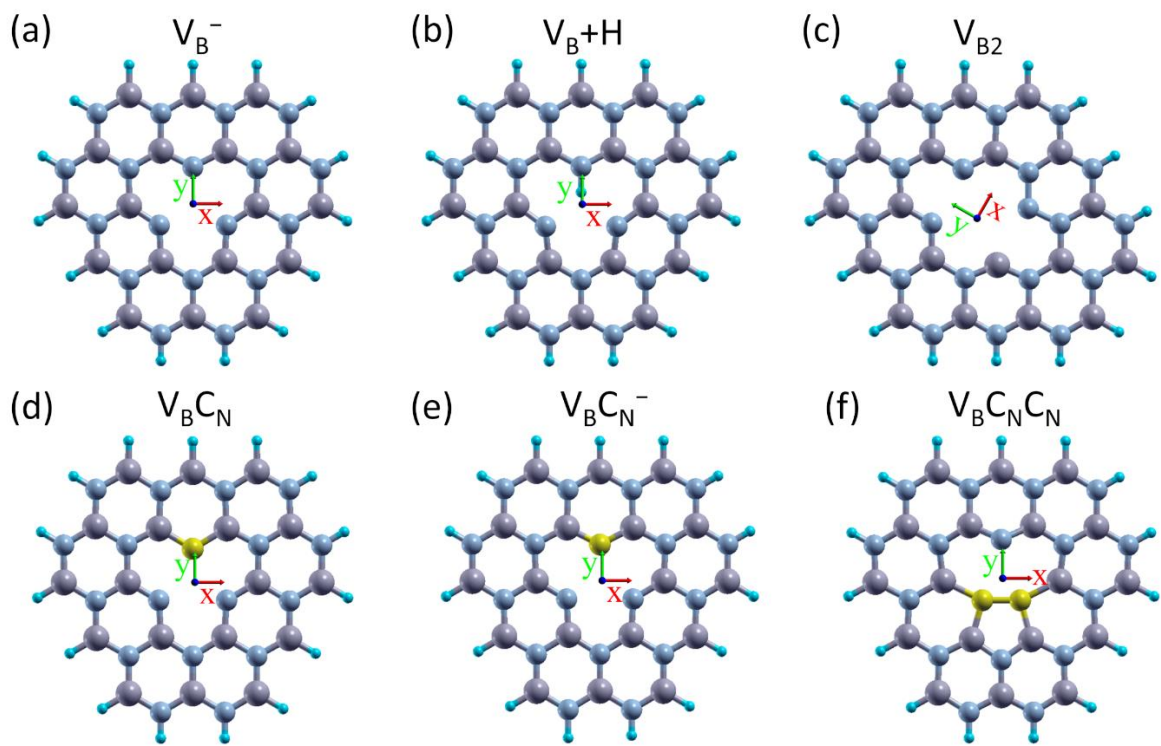

**Figure S1.** Cluster models used in density matrix renormalization group (DMRG) calculations, with edge passivated by hydrogen atoms and including two layer of rings around the defect centres. Coordinate axes marked out in each defect centre are the principle axes of the calculated EPR *g*-tensor as presented in Table 2 in the main text, here we use right-handed coordinate system. (a) $V_B^-$; (b) $V_B$+H; (c) $V_{B2}$; (d) $V_BC_N$; (e) $V_BC_N^-$; (f) $V_BC_NC_N$. Note that in $V_BC_NC_N$ the dangling nitrogen atom is tilted out-of-plane. Gray balls are boron, blue balls are nitrogen, green ball is hydrogen, while yellow balls are carbon.

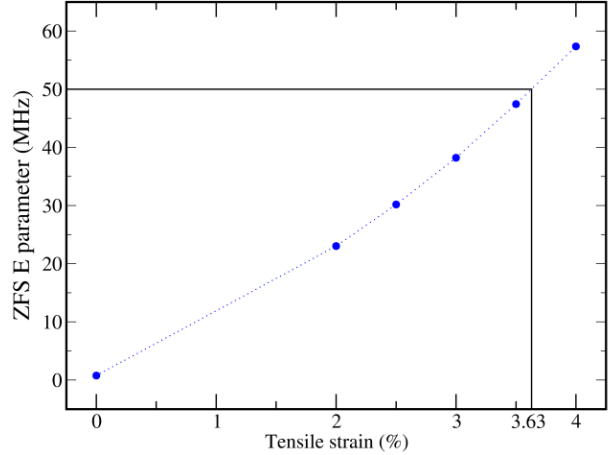

**Figure S2.** Strain tuning of the anisotropic zero field splitting(ZFS) E parameter of $V_B^-$ the defect center. The experimentally observed 50 MHz corresponds to a uniaxial tensile strain of 3.63%.

**Table S1.** Bright exciton state energies of boron vacancy-derived defects in hBN from GW-BSE calculations in the more complete energy spectral range, in comparison with DMRG GS geometry vertical excited state (ES) energies from the 2-layer ring cluster models of the corresponding defect center. The size of the active space used in DMRG calculations are given in the last column.

| Defect type | BSE exciton energies (eV) | DMRG ES energies (eV) | DMRG active space (#orbital, #electron) |
|---|---|---|---|
| $V_B^-$ | 3.89, 3.92, 3.96, 4.18, 5.10 | 2.17, 2.18, 2.91, 4.42, 4.44, 4.48 | (30, 10) |
| $V_B$+H | 0.70, 1.24, 1.61, 1.87, 2.10, 2.49 | 1.01, 1.11, 1.50, 1.93, 2.10, 2.65, 2.71 | (30, 10) |
| $V_{B2}$ | 2.86, 3.20, 3.84, 4.10, 4.60, 4.83 | 1.37, 3.04, 3.75, 4.92 | (30, 14) |
| $V_B C_N$ | 1.22, 1.35, 1.78, 2.00, 2.18, 2.22, 2.56, 2.77, 2.80, 4.05 | 2.24, 2.80, 3.12, 4.04, 4.06, 4.18 | (30, 10) |
| $V_B C_N^-$ | 2.42, 4.11, 4.26, 4.37, 4.52, 4.70 | 1.95, 1.98, 2.55, 5.06 | (30, 11) |
| $V_B C_N\,C_N$ | 1.43, 2.21, 2.35, 2.71, 2,78, 2.83, 2.91, 3.00, 3.20, 3.37, 3.45, 3.60, 3.65 | 1.05, 1.30, 2.20, 3.27, 3.30, 3.36 | (30, 10) |

**Table S2.** MR parameters of $V_B^-$ in comparison with previous literature.

| Triplet ground state EPR parameters of $V_B^-$ | | | | | |
|---|---|---|---|---|---|
| $A_{xx}$ (MHz) | $A_{yy}$ (MHz) | $A_{zz}$ (MHz) | D/h (GHz) | E/h (MHz) | Memo. |
| 47.5 | 46.0 | 88.1 | 3.3 | 0.8 | This work. |
| 47 | 47 | n.a. | 3.48 | 50 | Nat. Mater. **19**:540; experiment. |
| 47.9 | 46.1 | 91.6 | 3.47 | 0 | NPJ Comp. Mater. **6**:41; theory. |
| 54.5 | 49.1 | 102.5 | 3.3 | 0 | Comm. Phys. **3**:153; theory. |

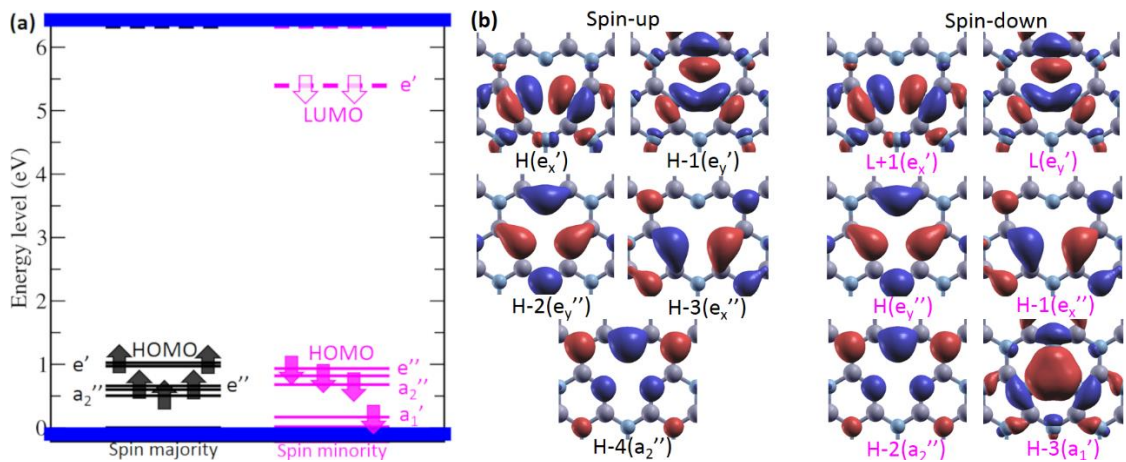

**Figure S3.** (a) Ground state GW electronic energy level diagram of the $V_B^-$ triplet defect at $\Gamma$ point, and (b) the corresponding electronic orbital wavefunction plots for in-gap levels.

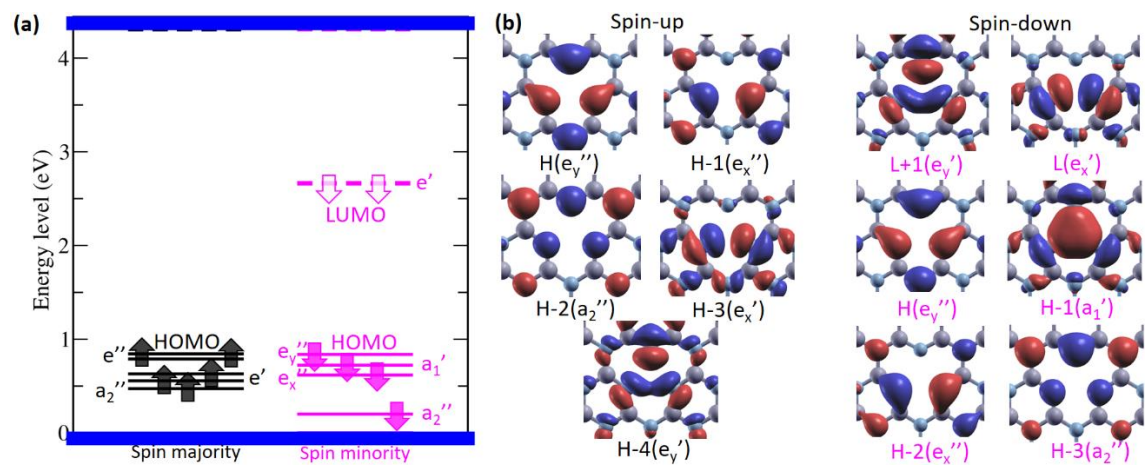

**Figure S4.** (a) Ground state DFT-HSE electronic energy level diagram of the $V_B^-$ triplet defect at $\Gamma$ point, and (b) the corresponding electronic orbital wavefunction plots for in-gap levels.

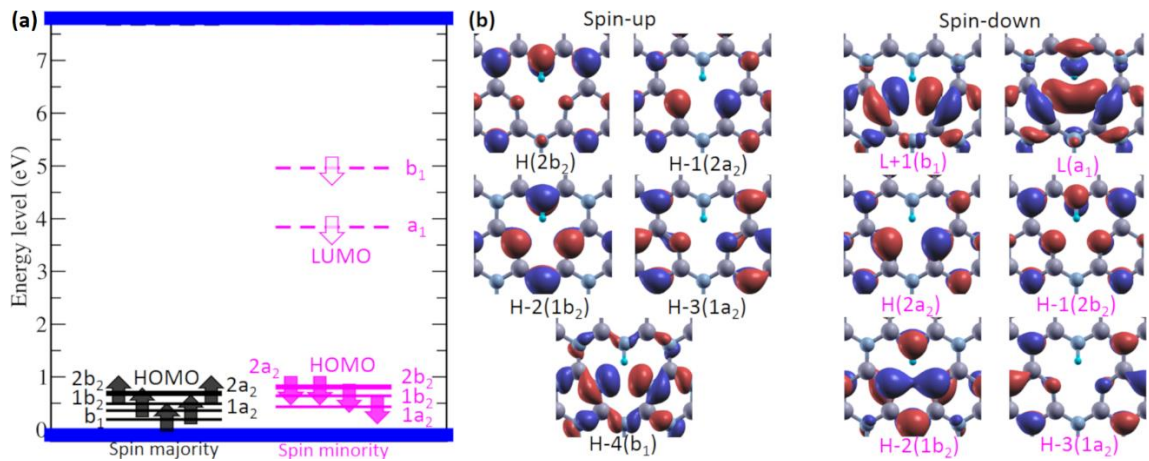

**Figure S5.** (a) Ground state GW electronic energy level diagram of the $V_B+H$ triplet defect at Γ point, and (b) the corresponding electronic orbital wavefunction plots for in-gap levels.

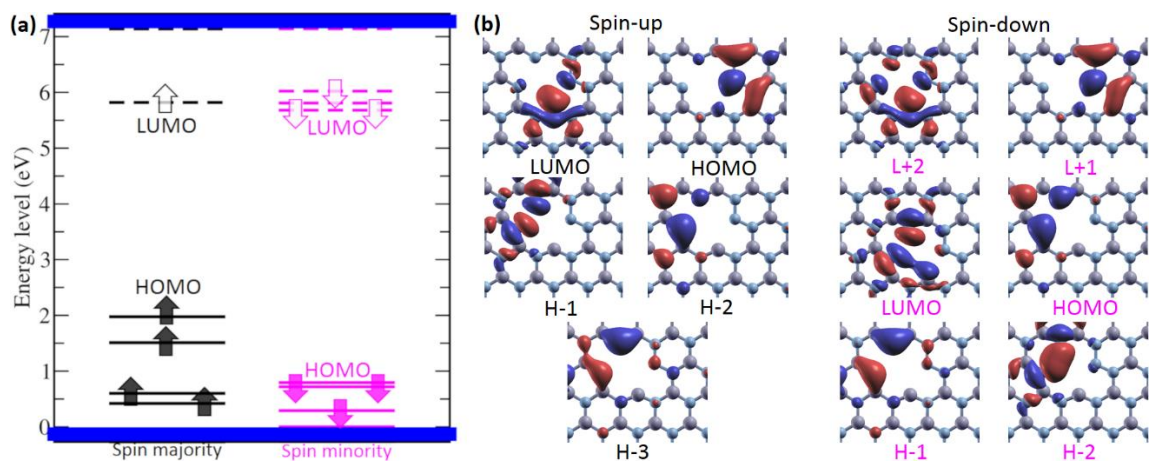

**Figure S6.** (a) Ground state GW electronic energy level diagram of the $V_{B2}$ triplet defect at Γ point, and (b) the corresponding electronic orbital wavefunction plots for in-gap levels.

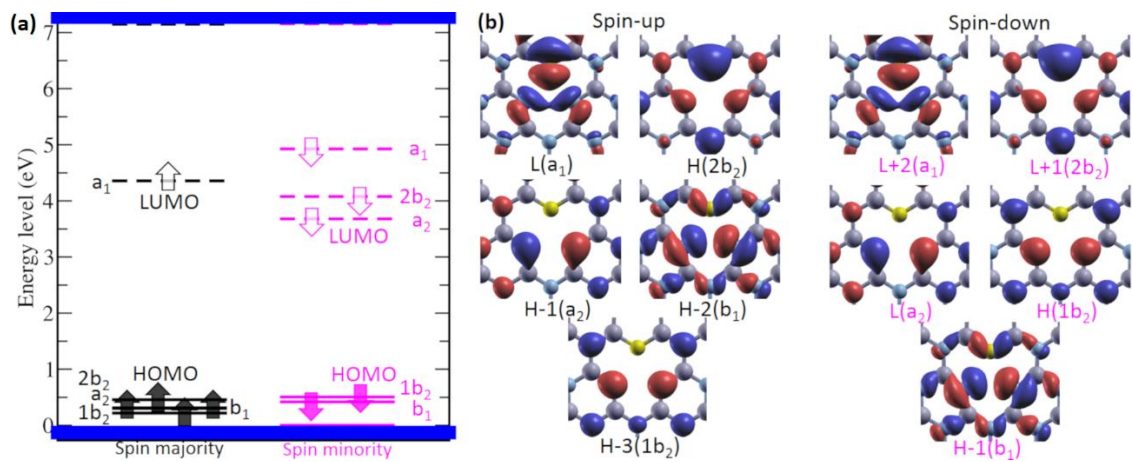

**Figure S7.** (a) Ground state GW electronic energy level diagram of the $V_BC_N$ triplet defect at Γ point, and (b) the corresponding electronic orbital wavefunction plots for in-gap levels.

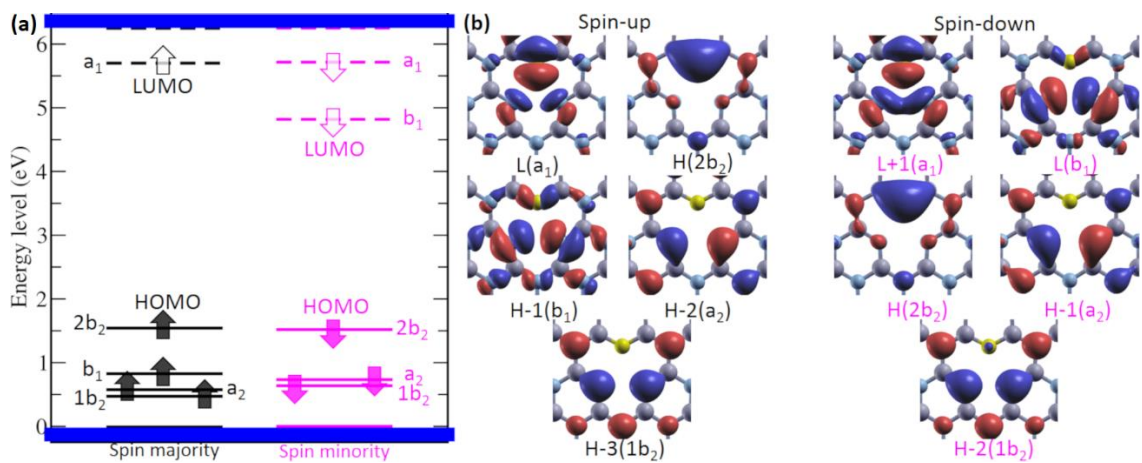

**Figure S8.** (a) Ground state GW electronic energy level diagram of the $V_BC_N^-$ doublet defect at Γ point, and (b) the corresponding electronic orbital wavefunction plots for in-gap levels.

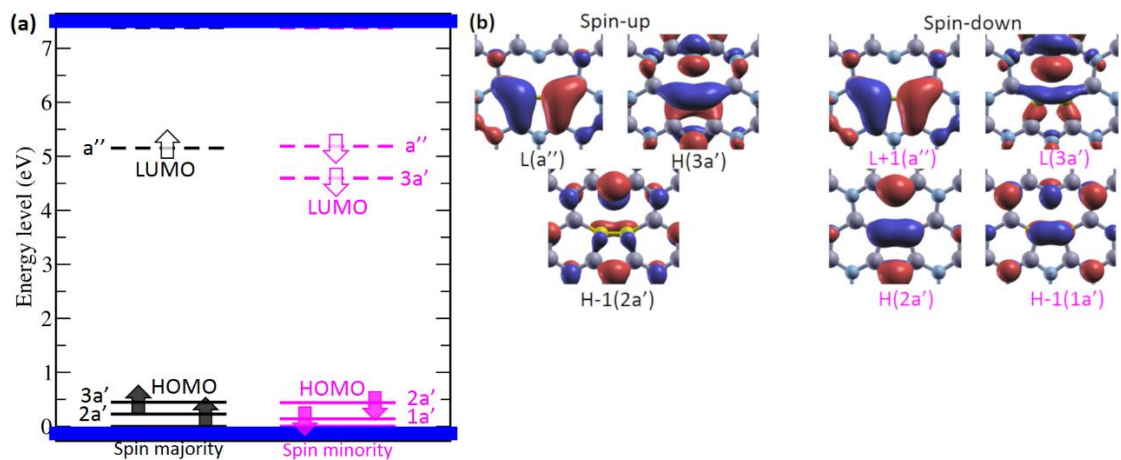

**Figure S9.** (a) Ground state GW electronic energy level diagram of the $V_BC_NC_N$ doublet defect at **Γ** point, and (b) the corresponding electronic orbital wavefunction plots for in-gap levels.